\documentstyle[epsf,rotate]{EuroPhys}

\newif\ifboo \boofalse

\begin{document}
\newcommand{\sharm}[2]{Y_{#1}^{(#2)}}
\newcommand{\pder}[2]{\frac{\partial #1}{\partial #2}}
\newcommand{\der}[2]{\frac{d #1}{d #2}}
\newcommand{\oforder}[1]{ {\cal O}(#1)}

\euro{}{}{}{}
\Date{}
\shorttitle{Transport Mean Free Path for Magneto-Transverse Light Diffusion }

\title{Transport Mean Free Path for Magneto-Transverse Light Diffusion }
\author{David Lacoste and Bart A. van Tiggelen }
\institute{Laboratoire de Physique et Mod\'elisation des Milieux Condens\'es\\
CNRS/Maison de Magist\`eres, Universit\'e Joseph Fourier \\
B.P. 166, 38042 Grenoble Cedex 9, France}

\rec{}{}

\pacs{
\Pacs{78}{20.Ls}{Magneto-optical effects}
\Pacs{42}{68.Ay}{Propagation, transmission, attenuation, and radiative transfer}
}

 \maketitle

\begin{abstract}
 We derive an expression for the transport mean 
 free path  $\ell^*_\perp$ associated with magneto-transverse
 light diffusion for a random collection of Faraday-active
 Mie scatterers. This expression relates the  
 magneto-transverse diffusion in multiple scattering directly to the magneto-transverse
 scattering of a single scatterer. 
\end{abstract}

Magneto-transverse light diffusion - more popularly
known as the ``Photonic Hall Effect" (PHE) - has been predicted theoretically
some three years ago \cite{prl}, and has been confirmed   experimentally  
one year later 
\cite{nature}. Phenomenologically, the effect has many
similarities to the well-known electronic Hall effect: given
a diffusion current $\mathbf{J}$, the presence of an 
external and constant magnetic field creates a flow
in the ``magneto-transverse" 
direction $q \mathbf{B} \times \mathbf{J}$, with $q$ 
the charge of the current carriers. By the non-existence of photon charge,
the physics of the PHE seems 
 different, and compares perhaps better to
the so-called Beenakker-Senftleben effect in
dilute gazes \cite{BS}. The evident driver behind the electronic Hall effect
is the Lorentz force acting on a charged particle while colliding with
the impurities. According to Ref.~\cite{prl} the 
PHE finds its origin in   the Faraday effect for 
dielectric  scatterers, that slightly changes their
 scattering amplitude.
 The charge $q$ is
  replaced by a material parameter $V$ with the symmetry of charge,
   quantifying
  the Faraday effect of the particle's material.
  In a homogeneous medium  the Faraday effect implies a
  rotation $VB$ per unit length  of the polarization vector of 
  linearly polarized light.
   Two other 
    magneto-optical effects in multiple scattering,
    such as the suppression of coherent backscattering 
    in a magnetic field \cite{sj,ralf}, and ``Photonic Magneto-Resistance"
    \cite{anjaprl} are known to exist, and basically originate from
    the same Faraday effect.

In an isotropic medium, Fick's phenomenological law relates the diffusion current to 
the energy-density gradient $\nabla \rho$ according to
$\mathbf{J} = -D_0 \mathbf{\nabla} \rho $ \cite{tom}. 
$D_0$ is the conventional diffusion constant 
for radiative transfer and is usually 
related to the transport mean free path $\ell^*$ and 
the transport velocity $v_E$ 
according to $D= \frac13 v_E \ell^*$. The velocity is relevant only for
dynamical experiments.  Fick's law applies
to media much larger than $\ell^*$ and, when supplied by boundary
conditions involving the incident flux, can be solved for the emerging current.
In a magnetic field,
the diffusion constant must be replaced by a second-rank tensor.
By Onsager's relation $D_{ij}(\mathbf{B}) =D_{ji}(\mathbf{-B})$, the part linear in the external magnetic field
must be an  antisymmetric tensor, and Fick's law becomes,
\begin{equation}
\mathbf{J} = - \mathbf{D}(\mathbf{B}) \mathbf{\cdot}\nabla \rho =  
-D_0\nabla \rho - D_\perp \mathbf{B} \times \nabla \rho\, .\label{symm}
\end{equation}
The    term containing $D_\perp$ 
describes a magneto-transverse diffusion current. In analogy to $D_0$,
 we shall define
the transport mean free path $\ell^*_\perp$ for magneto-transverse diffusion
as $D_\perp= \frac13 v_E \ell^*_\perp$.  For the
electronic Hall effect,  $\ell^*_\perp$
would be  proportional  to the Hall conductivity  $\sigma_{xy}$, whose sign 
is determined by the charge of the current carriers.  
Similarly,   $\ell^*_\perp$ of the PHE  can have both
 signs depending on the scattering.

Anisotropy in the scattering cross-section - quantified by
the familiar anisotropy factor $\left<\cos\theta\right>$ \cite{henk} - 
 is  well known to make
the transport mean free path $\ell^*$ different from the extinction length 
$\ell$ \cite{da}.
 The latter is the average distance between two subsequent scattering events.
In this Letter we will show that the PHE can be understood as
a generalization of this anisotropy factor to magneto light diffusion,
which establishes  a difference in scattering between ``upward" and ``downward" 
directions (with respect to the plane of incident light and magnetic field). To this end we use our solution
 for the Faraday-active dielectric sphere \cite{josa} to relate
 the PHE in multiple scattering, quantified by 
 $\ell^*_\perp$, directly to the PHE of one single particle. 
Although such a link may be physically clear, it is not evident
from previous work \cite{prl}.  A microscopic
approach provides both the exact sign as well as the role
of anisotropy, which will enable us 
to conclude this Letter with a  realistic comparison to
reported experiments \cite{nature}.

The magneto-active dielectric sphere has been discussed
by Ford et al. \cite{ford}. Experiments and symmetry arguments
show the PHE to be linear in $B$.   
Therefore we have 
derived a linear perturbation formula for 
the scattering amplitude $\mathbf{T}_{\mathbf{pp'}}
(\mathbf{B})$  
of one magneto-active dielectric sphere \cite{josa}.  
This amplitude relates the scattered field in direction ${\mathbf p}'$ to the
incoming plane wave with wave vector $\mathbf{p}$. 
The differential cross-section, proportional to
the modulus squared of the scattering amplitude, 
must satisfy the reciprocity relation $\mathrm{d}\sigma/ \mathrm{d}
\Omega (\mathbf{p}\rightarrow \mathbf{p'}, \mathbf{B} )
= \mathrm{d}\sigma /\mathrm{d}
\Omega (\mathbf{-p'}\rightarrow \mathbf{-p}, \mathbf{-B} )$. A
magneto-cross-section proportional to  
$(\widehat{\mathbf{p}} \cdot \widehat{\mathbf{B}})$ or
$(\widehat{\mathbf{p}}' \cdot \widehat{\mathbf{B}})$ 
is parity-forbidden since $\mathbf{B}$ is a pseudo-vector and
$\mathbf{p}$ a vector. Together
with the rotational symmetry of the sphere it must have the form,
\begin{equation}
{1\over \sigma_{\mathrm{tot}}}{\mathrm{d}\sigma \over \mathrm{d}
\Omega} (\mathbf{p}\rightarrow \mathbf{p'}, \mathbf{B} )
=  F_0(\theta) + 
\mathrm{det}(\widehat{\mathbf{p}}, 
\widehat{\mathbf{p}}', \widehat{\mathbf{B}}) \, F_1(\theta)  
 \, ,
\label{mmie}
\end{equation} 
where $\cos\theta = \widehat{\mathbf{p}}\cdot \widehat{\mathbf{p}}'$ defines
the scattering angle $\theta$, 
$\sigma_{\mathrm{tot}}$ is the total cross-section and 
$\mathrm{det}(\mathbf{A}, 
 {\mathbf{B}}, {\mathbf{C}})=  {\mathbf{A}}
\cdot ( {\mathbf{B}} \times  {\mathbf{C}})$ the scalar
determinant that can be constructed from  three vectors. 
The second term in Eq.~(\ref{mmie}) will be called the magneto cross-section. 
For a small, Rayleigh scatterer one
 finds 
 $F_1(\theta)\sim VB \cos\theta c_0/\omega $.
The antisymmetry of this magneto cross-section between 
forward scattering and backscattering causes the PHE to vanish.  
The Mie solution  breaks this symmetry and a PHE was seen
to emerge \cite{josa}.

We will use  field techniques developed in Refs.~\cite{da,nema} to calculate
the magneto-transverse transport mean free path for 
a random collection of identical Faraday-active dielectric spheres.
The four-rank tensor $L_{ijkl, \mathbf{p}\mathbf{p'}}( 
\mathbf{q}) $ linearly connects  field correlations  
$\left<E_l(\omega ,\mathbf{p}+
\mathbf{q}/2)\right.$ 
$ \left. 
 \bar{E}_j (\omega,\mathbf{p}-\mathbf{q}/2) \right>$ of incident and
outgoing fields in space; $\mathbf{q}$ is the Fourier variable of space and
$\mathbf{p}$ is the optical wave vector. On long  length scales ($\mathbf{q}
 \rightarrow 0$) and without absorption it takes the diffuse form,
 \begin{equation}
 L_{ijkl, \mathbf{p}\mathbf{p'}}( \mathbf{q}, \mathbf{B}) =
{ l_{ik}(\mathbf{p},\mathbf{q},\mathbf{B})  {l}_{lj}(
-\mathbf{p'},-\mathbf{q},-\mathbf{B}) 
\over
  \mathbf{q\cdot D}(\mathbf{B}) \mathbf{\cdot q}}
\, .
\label{ijkl}
\end{equation}
The symmetric form of the numerator is imposed by the
reciprocity principle. Rigorous transport
theory yields \cite{nema},
\begin{equation}
\mathbf{l}(\mathbf{p},\mathbf{q},\mathbf{B}) = \mathrm i
 \left[ \mathbf{G}(\mathbf{p},\mathbf{B}) -\mathbf{G}^*(\mathbf{p},\mathbf{B})
 \right]  
-\mathrm i \mathbf{G}(\mathbf{p},\mathbf{B})
\cdot \mathbf{\Gamma}(\mathbf{p},\mathbf{q},\mathbf{B} )\cdot
 \mathbf{G}^*(\mathbf{p},\mathbf{B})
\label{gamma1}
\, .
\end{equation}
$\mathbf{G}(\mathbf{p},\mathbf{B} )$
denotes the Dyson Green's tensor of the ensemble-averaged
 electric field to be specified later;
  the asterisk denotes hermitean conjugation
 in polarization space. We left out explicit reference to the 
 optical frequency $\omega$. 
 
 The tensor 
 $\mathbf{\Gamma}(\mathbf{p},\mathbf{q},\mathbf{B} )$ is linear in
 $\mathbf{q}$. In real space,
 the wavenumber $\mathbf{q}$ corresponds to the gradient
 in Fick's law (\ref{symm}). The exact relation between  $ \mathbf{\Gamma}$ and the diffusion tensor is
 \cite{prl,nema}, 
\begin{eqnarray}
 \mathbf{D}(\mathbf{B}) \cdot \mathbf{q} = v_E
{\pi c_0 \over \omega^2} \sum_{\mathbf{p}} \, \mathbf{p} \, \mathrm{Tr}\,
\mathbf{G}(\mathbf{p},\mathbf{B})\cdot \mathbf{\Gamma}(
\mathbf{p},\mathbf{q}, \mathbf{B}) \cdot  \mathbf{G}^*
(\mathbf{p},\mathbf{B}) 
=  
v_E \ell \int {\mathrm{d}^2\widehat{\mathbf{p}}\over 4\pi} \, \widehat{\mathbf{p}}
\, \mathrm{Tr}\, 
\mathbf{\Gamma}(\mathbf{\widehat{p}},\mathbf{q}, \mathbf{B})\, , 
 \label{trgamma}\end{eqnarray}
This shows that only the trace of $\mathbf{\Gamma}$
comes in. The term $\mathrm{Tr}\, \delta \mathbf{\Gamma}$
linear in $\mathbf{B}$ provides the PHE.

 For a low density $n$ of
 particles   with
 T-matrix $\mathbf{T_{pp'}}(\mathbf{B})$, $\mathbf{\Gamma}$
 obeys the
 Bethe-Salpeter equation \cite{nema},
 \begin{equation}
 \mathbf{\Gamma}(\mathbf{p},\mathbf{q},\mathbf{B} )
 = 2(\mathbf{p\cdot q})  
 + n \sum_{\mathbf{p'}} \mathbf{T_{pp'}}(\mathbf{B})  \cdot \mathbf{G}
 (\mathbf{p'}, \mathbf{B}) \cdot \mathbf{\Gamma}(\mathbf{p'},\mathbf{q},
 \mathbf{B} )
 \cdot \mathbf{G}^*(\mathbf{p'}, \mathbf{B}) \cdot 
  \mathbf{T}^*_{\mathbf{pp'}}(\mathbf{B})
\, . \label{BS}
\end{equation}
We shall solve this equation for the magneto-active Mie particle,
up to linear contributions in $\mathbf{B}$.  
$\mathbf{\Gamma}$ determines 
the anisotropy in scattering;
without magnetic field it can be ascertained 
that $\mathbf{\Gamma}^0(\mathbf{p},\mathbf{q}) =  
2(\mathbf{p\cdot q})/(1-\left<\cos\theta\right>)$, 
with $\left<\cos\theta\right>$
the  anisotropy factor in scattering  
 \cite{henk}.

Eq.~(\ref{BS})
shows that $\mathbf{\Gamma}$ is a hermitean tensor, linear in $\mathbf{q}$.
We will ignore  the longitudinal field  in Eq.~(\ref{BS})
 that turned
out to be of higher order.   
This makes $\mathbf{\Gamma}$ as well as $\mathbf{G}$  effectively a
hermitean $2\times 2$ matrix.  
It is convenient to express all such matrices with respect to  a helicity 
base, where $\sigma(\mathbf{p})= -\sigma(\mathbf{-p})=\pm 1$ 
the helicity
of a plane wave with wave vector $\mathbf{p}$. They must be
a linear combination
of the identity $U$ and the three Pauli spin matrices 
$\mathbf{\sigma}_{x,y,z}$ \cite{quantum}. If the $z$-axis is taken along
the wave vector $\mathbf{p}$, the Green's tensor can be written as
 $ G_{\sigma\sigma'}(\mathbf{p,B}) = G_0(p) \, {U}
+ G_1(p)\, (\mathbf{B\cdot p})\, \mathbf{\sigma}_z$,
with $G_0(p) = 1/[\omega^2/c_0^2 -p^2 + \mathrm i \omega/c_0\ell] $ the
  Dyson Green's function in terms of the
extinction length  $\ell$; furthermore $G_1(p) \sim G_0^2$. 

We separate the
magneto-optical term as $\mathbf{\Gamma}(\mathbf{B})
 = \mathbf{\Gamma}^0 + \delta\mathbf{\Gamma}(\mathbf{B}) $.
Mirror-symmetry imposes that $\mathbf{T}_{\mathbf{pp'}}(\mathbf{B})
= \mathbf{T}_{\mathbf{-p-p'}}(\mathbf{B})$ so that, by Eq.~(\ref{BS}),
$\mathbf{\Gamma}(\mathbf{p,q,B}) = -\mathbf{\Gamma}(\mathbf{-p,q,B})$.
Finally, this implies
\begin{eqnarray}
\delta {\Gamma}_{\sigma\sigma'}(\mathbf{p},\mathbf{q}, \mathbf{\widehat{B}})
=   a_1 (\mathbf{p\cdot   q\times \widehat{B}}) \, {U} &+&  
\mathbf{\sigma}_z \left[ a_2 (\mathbf{\widehat{B} \cdot q}) + a_3   (\mathbf{\widehat{B} \cdot p})
(\mathbf{p\cdot q}) \right]\nonumber \\
&+& a_4 \left[ (\widehat{B}_yq_x +\widehat{B}_xq_y)\, \mathbf{\sigma}_x +  
(\widehat{B}_yq_y -\widehat{B}_xq_x)\, \mathbf{\sigma}_y \right]   
\label{ansatz}
\end{eqnarray}
in terms of 4   
real-valued coefficients $a_n$ to be determined. They correspond to the
4 Stokes parameters $I$, $Q$, $U$ and $V$ of the diffuse light. 
The presence of matrices other than the unit matrix 
 in Eq.~(\ref{ansatz})
implies that the diffuse radiation (\ref{gamma1}) becomes
 polarized in the presence of a magnetic field.
 By Eq.~(\ref{trgamma}), the PHE  is determined  by $\mathrm{Tr}\, \delta \mathbf{\Gamma} \sim a_1$, 
 for which Eq.~(\ref{BS}) provides a closed
 equation, as we will show now.

  The linearization of Eq.~(\ref{BS}) in the magnetic field
 generates three coupled contributions: $\delta \mathbf{\Gamma}=
 \delta \mathbf{\Gamma}_1+\delta \mathbf{\Gamma}_2+\delta \mathbf{\Gamma}_3$.
The first comes   from
$ \delta \mathbf{\Gamma}(\mathbf{B})$ itself, the second
 from $\delta \mathbf{T_{pp'}}
(\mathbf{B})$ and
the third from $\delta \mathbf{G}(\mathbf{B})$.  
The 
$T$-matrix of a Mie sphere without magnetic field depends on
the scattering angle $\theta$ and the azimuthal angle $\phi$ in a frame
where 
 the incident wave vector is directed along the $z$-axis\cite{henk}. With respect to  a helicity 
base, it takes the form,
 \begin{equation}
 {T}_{\sigma \mathbf{p}\sigma'\mathbf{ p'}}^0 = -{2\pi \mathrm i\over p}
 \mathrm e ^{-\mathrm i \sigma\phi} \left[ 
\bar{S}_1(\theta) + 
\bar{S}_2(\theta)\sigma(\mathbf{p})\sigma'(\mathbf{p'}) \right] = 
-{2\pi \mathrm i\over p} \mathrm e ^{-\mathrm i \sigma\phi}\left[ 
(\bar{S}_1 + 
\bar{S}_2 ) \, U + (\bar{S}_1- 
\bar{S}_2 ) \sigma_x \right] \, . \label{mie0}
\end{equation}
The bars denote complex conjugation and stem from our different
convention than the one used in Ref.~\cite{henk}.
Since $\sigma_x \sigma_y = \mathrm i\sigma_z $ and $\sigma_z \sigma_x = \mathrm 
i\sigma_y $
many terms generated by  $\delta\mathbf{\Gamma} $ in Eq.~(\ref{BS}) contain
one of the traceless  Pauli spin matrices. Since $\delta G \sim \sigma_z$ 
it follows that $\mathrm{Tr}\, \delta \mathbf \Gamma_3=0$.
Since $\sigma_x^2= U$ one term involving $a_4$ survives
in $\mathrm{Tr}\, \delta\mathbf \Gamma_1$ but is 
eliminated by the  integral over the azimuthal angle $\phi$. 
The first contribution simplifies to,

\begin{eqnarray}
\mathrm{Tr}\, \delta \mathbf{\Gamma}_1(\mathbf{p,q,B}) =  
 \int \mathrm{d}^2\widehat{\mathbf{p}}'\, F_0(\theta) 
 \mathrm{Tr}\, \delta \mathbf{\Gamma}(\mathbf{p',q,B})= 
2a_1\left<\cos\theta \right>
 (\mathbf{p \cdot q \times \widehat{B}})\, .
 \label{een} 
 \end{eqnarray}
with $F_0(\theta)$ defined in Eq.~(\ref{mmie}). 
 The second equality follows
by substituting   $\mathrm{Tr}\, \delta \mathbf{\Gamma}(\mathbf{p',B,q}) = 
2a_1 
(\mathbf{p'  \cdot q \times \widehat{B}})$.  
The   contribution $\mathrm{Tr}\, \delta\mathbf{\Gamma}_2$ is obtained by
substituting $\mathbf{\Gamma}^0
=2 (\mathbf{p\cdot q})/(1-\left< \cos\theta \right> )$,
  
\begin{equation}
\mathrm{Tr}\, \delta \mathbf{\Gamma}_2(\mathbf{p,q,B}) =  
 {2\over 1-\left< \cos\theta \right>} \int \mathrm{d}^2\widehat{\mathbf{p}}'\, 
 (\mathbf{p'\cdot q})\, \mathrm{det}(\mathbf{\widehat{p},\widehat{p}',\widehat{B}}) F_1(\theta)\, , 
 \label{twee} 
 \end{equation}
and involves  the
magneto cross-section
  introduced in Eq.~(\ref{mmie}). From $\mathrm{Tr}\, \delta \mathbf{\Gamma}_3=0$,
  and Eq.~(\ref{ansatz}) we have
$\mathrm{Tr}\, \delta \mathbf{\Gamma}_1 = 2a_1(\mathbf{p \cdot q\times
 \widehat{B}}) 
- \mathrm{Tr}\, \delta \mathbf{\Gamma}_2$.  Eqs.~(\ref{een}) 
 and (\ref{twee}) hold for any $\mathbf{q}$ and it is convenient to choose
  $\mathbf{q} =   \mathbf{\widehat{p}\times \widehat{B}}$.
  This yields the desired equation,
  
\begin{eqnarray}
a_1 = {1\over (1-\left< \cos\theta \right>)^2} 
\int \mathrm{d}^2\widehat{\mathbf{p}}'\, 
  {\mathrm{det}^2(\mathbf{\widehat{p},\widehat{p}',\widehat{B}}) 
 \over |\mathbf{\widehat{p}} \times \mathbf{\widehat{B}}|^2 }\,  F_1(\theta)
= {\pi\over (1-\left< \cos\theta \right>)^2}
  \int_0^\pi \mathrm{d} \theta \sin^3 \theta F_1(\theta)\, .
\label{a1}
\end{eqnarray}
The last equality follows upon integration over the azimuthal angle 
$\phi$.

We can now insert our result for $\mathrm{Tr}\, \mathbf{\Gamma}$
into formula~(\ref{trgamma}) for the diffusion tensor.
Since $\mathrm{Tr}\, \delta\mathbf{\Gamma}= 2a_1p_i\epsilon_{ijk}q_j
\widehat{B}_k$, the
final result reads,
 \begin{equation}
 {D}_{ij}(\mathbf{B}) = \frac13 v_E {\ell \over 1-\left< \cos\theta\right>} \,  
 \delta_{ij} +
 \frac13 v_E a_1 \ell\,  {\epsilon_{ijk} \widehat{B}_k}
 \, .\label{final}
 \end{equation}  
This equation identifies $\ell^*= \ell/(1-\left< \cos\theta\right>)$ 
as the transport mean free path of light in multiple scattering, and
 $\ell_\perp^*\equiv a_1\ell$ 
as the transport mean free path
for magneto-transverse light scattering. In Ref.~\cite{josa} we have shown that the 
integral in Eq.~(\ref{a1}) equals exactly  the normalized 
PHE of one magneto-active Mie sphere. 
Eq.~(\ref{final}) thus states that the PHE in multiple scattering
is directly proportional to the normalized PHE of one single Mie sphere, 
including the same sign, and amplified by the factor $1/(1-\left<\cos\theta
\right>)^2$. Note that  the magneto-transverse
transport mean free depends  even more
on the anisotropy factor $\left<\cos\theta\right>$ than the conventional
transport mean free path $\ell^*$. Unfortunately, we have no simple explanation
for this.

\begin{figure}
  \hfil
  \begin{minipage}[b]{6.cm}
    \epsfxsize 6.cm
    \rotate[r]{ \epsfbox{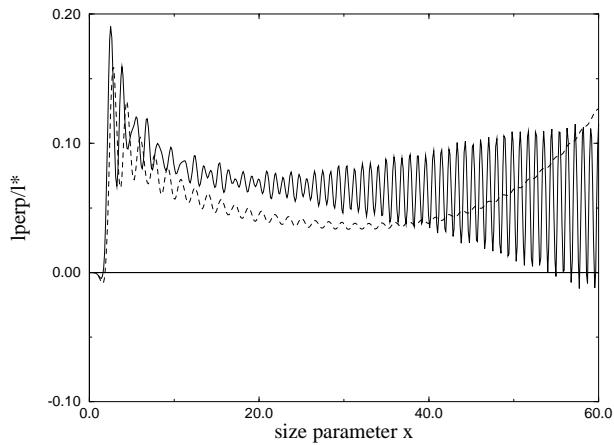} }
  \end{minipage}
\caption
{ \label{plot}
   Ratio of  magneto-transverse transport
mean free path and isotropic transport mean free path 
$\ell_\perp^*/\ell^* = (1-\left< \cos\theta \right>)a_1$,
as a function of the size parameter
$x=2\pi a/\lambda$ of the Mie spheres. The index of refraction
  $m=1.128$ of the spheres corresponds to CeF$_3$ in glycerol.
The ripple structure is attributed
to resonances in the various partial waves
that built up the scattering amplitude for large $x$.
For size parameters $x<1$ (Rayleigh regime) we find that  
$\ell_\perp/\ell^*\sim x^5$ vanishes
rapidly. 
Dashed line: the same for $m \rightarrow 1 $ (Rayleigh-Gans regime), for which a finite
value is seen to survive.
}
\end{figure}

In Fig. 1 we show
  $\ell_\perp^*/\ell^*$ as a function of the size parameter,
  for an index of refraction $m=1.128$, corresponding to 
 CeF$_3$ in glycerol.  Around $x\approx 40 $, (radius $2\, \mu$m) we calculate
  $\ell_\perp^*/\ell^* =
  + 0.06 \, V B\lambda$ which, for $V=-1100 $ rad/mT (at temperature
  $T=77$ K)  and vacuum wavelength 
 $\lambda_0 = 0.457 \, \mu$m  yields  
 $\ell_\perp^*/\ell^* = - 2 \cdot 10^{-5} $/T.   
The experimental value is $\ell_\perp/\ell^* \approx - 1.1 \pm 0.3 
\cdot 10^{-5} /\mathrm{T}$ 
 for  a 10 vol-$\%$ suspension \cite{nature}.
 We estimate a systematic error
 of at least a factor of two in {\it attributing}  values to
 $\ell_\perp $ and $\ell^*$ on the basis of Fick's law.
 Another uncertain factor is the broad size distribution
 in this sample, which probably washes out
  the oscillations of $\ell_\perp^*/\ell^*$ as
 a function of size parameter $x$ that are seen
 in Fig.~1. 
Besides CeF$_3$, the
present theory is also able to reproduce the measured sign and  magnitude
for the PHE of polydisperse samples containing ZnS, Al$_2$O$_3$,  TiO$_2$
and EuF$_2$. The sign can change as a function of $x$ and $m$
and cannot be predicted by a simple argument known to us. 
The estimated anisotropy factor for our sample
 equals $\left<\cos\theta\right>\approx 0.9$.
 The amplification
 factor $1/(1-\left<\cos\theta
\right>)^2$ for the PHE is therefore significant, and improves
 theoretical predictions made for
Rayleigh scatterers considerably \cite{nature}.

We conclude that the present theory for spheres of arbitrary size
 is a significant step forward 
in the qualitative and quantitative understanding of  
 the magneto-transverse light diffusion.
We thank  Geert Rikken, Anja Sparenberg and Venkatesh Gopal for 
the many discussions. This work has been made 
possible by the Groupement de Recherches POAN.

\end{document}